\documentclass[aps,prb,twocolumn,superscriptaddress,showpacs]{revtex4}
\usepackage{graphicx,bm,color,float,textcomp}

\newcommand\mH{magnetization}
\newcommand\Hlll{$H \! \parallel \! [111]$}

\newcommand\Hloo{$H \! \parallel \! [100]$}
\newcommand\GTO{$\rm Gd_2Ti_2O_7$}
\newcommand\PRB[3]{Phys. Rev. B {\bf {#1}}, {#2} {(#3)}}
\newcommand\PCM[3]{J. Phys.: Condens. Matter {\bf {#1}}, {#2} {(#3)}}

\newcommand\PPRL[3]{Phys. Rev. Lett. {\bf {#1}}, {#2} {(#3)}}
\newcommand\ibid[3]{{\it ibid.} {\bf {#1}}, {#2} ({#3})}
\newcommand\Tno{$T_{N1}$}
\newcommand\Tnt{$T_{N2}$}
\newcommand\Mlong{$M_\parallel$}
\newcommand\MlongH{$M_\parallel (H)$}
\newcommand\Mtran{$M_\perp$}

\begin{document}
\title{Novel magnetic phases in a Gd$_2$Ti$_2$O$_7$ pyrochlore for a field applied along the $[100]$ axis}
\author{O.A.~Petrenko}
\author{M.R.~Lees}
\author{G.~Balakrishnan}
\affiliation{Department of Physics, University of Warwick, Coventry CV4 7AL, United Kingdom}
\author{V.N.~Glazkov}
\author{S.S.~Sosin}
\affiliation{P.L.~Kapitza Institute for Physical Problems RAS, 119334 Moscow, Russia}
\begin{abstract}
We report on longitudinal and transverse \mH\ measurements performed on single crystal samples of \GTO\ for a magnetic field applied along the [100] direction.
The measurements reveal the presence of previously unreported phases in fields below 10~kOe in an addition to the higher-field-induced phases that are also seen for \Hlll, [110], and [112].
The proposed  $H$-$T$ phase diagram for the $[100]$ direction looks distinctly different from all the other directions studied previously.
\end{abstract}
\maketitle
The model of a Heisenberg antiferromagnet on a pyrochlore lattice has been the focus of theoretical attention for a number of years.
The highly degenerate ground state manifold for a system of spins on corner-shared tetrahedra interacting through nearest-neighbour exchange prevents a magnetic ordering both in the quantum and the classical
limits.\cite{Heis_theory1}
Further interactions (e.g. dipolar) can remove this degeneracy and stabilize a particular ordered structure.\cite{Heis_theory3}
To date, only two experimental realizations of a Heisenberg pyrochlore lattice antiferromagnet are known: \GTO\ (GTO) and {$\rm Gd_2Sn_2O_7$} (GSO).
The spin-orbit coupling is strongly reduced for a Gd$^{3+}$ magnetic ion since its electronic ground state is $^8S_{7/2}$ with $L=0$.
On cooling below 1~K, GTO and GSO develop different types of magnetic order\cite{Raju_PRB_1999,Champion_PRB_2001,Bertin_EPJB_2002,Bonville_JPCM_2003} which have been intensively studied over the last decade (for a recent review see Ref.~\onlinecite{Gardner_review_2010}).

The application of neutron scattering techniques to the determination of the magnetic structure in the ordered phases of GTO and GSO is hindered by the high neutron absorption cross section of naturally occurring Gd.
The zero-field $k=(\frac{1}{2}\frac{1}{2}\frac{1}{2})$ magnetic structure of \GTO\ is generally believed to be a multi-$k$ structure suggested by Stewart {\it et al.,}\cite{Stewart_JPCM_2004} however, this was recently challenged by Brammal {\it et al.}\cite{Brammal_PRB_2011,Stewart_PRB_comment_2012}.
While the zero-field structure in GSO appears to be simpler,\cite{Wills_JPCM_2006} its behaviour in field remains largely unexplored because of a lack of single crystal samples.

In this Rapid Communication we revisit the unusual $H$-$T$ phase diagram of GTO.
Initial interest in the field-induced behavior of GTO was driven by the heat capacity and susceptibility data obtained on a polycrystalline sample.\cite{Ramirez_PRL_2002}
Later work on single crystals showed that despite a nominal spin-only state, the magnetic properties of GTO are anisotropic and that the $H$-$T$ phase diagram contains three different ordered phases.~\cite{Petrenko_PRB_2004} 
The sequence of phase transitions in GTO in an applied magnetic field has been studied further using several techniques including muon spin relaxation,\cite{Dunsiger_PRB_2006} transverse \mH\ measurements~\cite{Glazkov_JPCM} and magnetic resonance.\cite{Sosin_PRB_2006}
Nevertheless, to date, there is no full theoretical description of the observed magnetic phases.

All previous studies of the magnetic phase transitions in GTO were limited to fields applied along the [111], [110], or [112] directions, while no data have been published for \Hloo.
Here we report on longitudinal and transverse \mH\  measurements performed on single crystals samples of \GTO\ for a magnetic field applied along the [100] direction.
These measurements reveal the presence of previously unreported field-induced phase transitions that occur in magnetic fields below $\sim10$~kOe, in addition to the transitions at higher applied fields that are also seen for \Hlll, [110], and [112].

Single crystal samples were prepared as described previously.\cite{Balakrishnan_JPCM_1998}
The principal axes of the samples were determined using the X-ray diffraction Laue technique; the crystals were aligned to within an accuracy of 2~degrees.
Two samples grown and aligned independently have shown no appreciable differences in their \mH\ behavior.

Longitudinal \mH, (\Mlong), measurements were made down to 0.5~K in applied magnetic fields of up to 70~kOe using a Quantum Design Magnetic Properties Measurement System SQuID magnetometer along with an i-Quantum
$^3$He insert.\cite{Shirakawa_JMMM_2004}
The \mH\ was measured both as a function of temperature in a constant magnetic field and as a function of applied field at constant temperature.
Because of the relatively large magnetic moments observed, de\mH\ effects had to be taken into consideration.

Transverse \mH, (\Mtran), measurements were performed using a homemade capacitance torquemeter mounted in a $^3$He cryostat with a base temperature of 0.4~K.
The sensor element was a flat capacitor formed by a rigid base and a flexible bronze cantilever with the sample attached.
The cantilever was aligned parallel to the external magnetic field.
The torque $\mathbf{T} = \mathbf{M_\perp} \times\mathbf{H}$, caused by the \mH\ component normal to the capacitor plates, leads to a bending of the cantilever and to a change in the sensor capacitance.
This technique has already been successfully applied to the study of the magnetic ordering in GTO.\cite{Glazkov_JPCM}
The sample for the torque measurements was cut in the shape of a thin plate of approximately $0.15\times 1\times 1$~mm$^3$, with its plane coinciding with a (110)-plane of the crystal.
It was glued on to the cantilever with a [001] axis parallel to the field and a [110] axis normal to the cantilever plane.
The field was applied in the sample plane, thus any de\mH\ effects were negligible.

\begin{figure}[tb]
    \begin{center}
    \includegraphics[width=0.95\columnwidth]{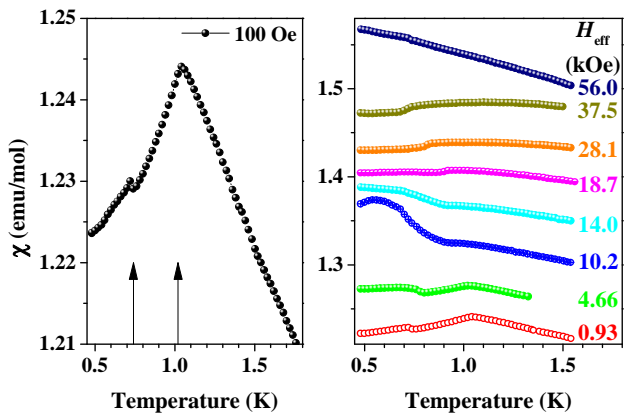}
    \end{center}
    \caption{\label{Fig1_M} (Colour online) Temperature dependence of the magnetic susceptibility of \GTO\ measured with a magnetic field applied along the  [100] direction.
    The left-hand panel shows the data collected in 100~Oe.
    The arrows indicate the two transition temperatures at $T_{N1}=1.02$~K and $T_{N2}=0.74$~K reported from the specific heat measurements in zero field.\cite{Petrenko_PRB_2004}
    The right-hand panel shows the temperature dependence of the susceptibility measured on warming in various higher fields.
    These curves are consecutively offset by 0.05~emu/mol for clarity.}
\end{figure}
The temperature dependence of the magnetic susceptibility in a magnetic field of 100~Oe applied along [100] is shown in the left-hand panel of Fig.~\ref{Fig1_M}.
Both the upper (\Tno) and lower (\Tnt) critical temperatures are clearly visible in the data.
In agreement with the previously reported data for \Hlll,\cite{Petrenko_JPCM_2011} the upper phase transition is observed at 1.05~K, a temperature marginally higher than $T_{N1}=1.02$~K found from the specific heat measurements,\cite{Petrenko_PRB_2004} while the lower transition temperature, 0.74~K, is identical for all the measurements.
The evolution of the temperature dependence of the magnetic susceptibility with applied field is shown in the right-hand panel of Fig.~\ref{Fig1_M}.

\begin{figure}[tb]
    \includegraphics[width=0.95\columnwidth]{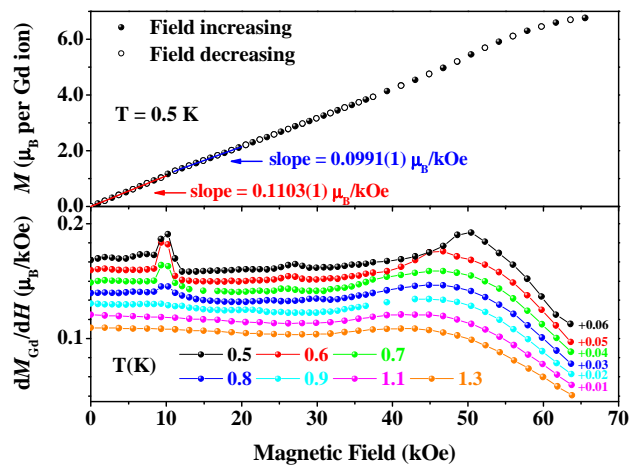}
    \caption{\label{Fig2_M} (Colour online) The upper panel shows the field dependence of \Mlong\ of \GTO\ at 0.5~K with \Hloo.
    The data taken for increasing and decreasing fields are almost indistinguishable.
    Linear fits to the data in the interval 0 to 10~kOe and 10 to 20~kOe reveal a 10\% change in the \MlongH\ slope at $H\approx 10$~kOe.
    The lower panel shows the field dependence of the $dM_\parallel/dH$ curves, obtained from the \MlongH\ data at different temperatures and then offset by the specified values.}
\end{figure}
The field dependence of the longitudinal \mH\ is shown in Fig.~\ref{Fig2_M} together with the $dM_\parallel/dH$ curves obtained from the \MlongH\ curves measured at different temperatures.
No appreciable hysteresis is observed in the \MlongH\ data.
In the highest applied field of 70~kOe, which translates into 65.5~kOe after taking into account the de\mH\ field, \Mlong\ is still growing at a considerable rate.
The maximum measured magnetic moment is 6.8$\mu_B$ per Gd ion, close to the value of 7$\mu_B$ expected for a state with $S=7/2$ and $L=0$.
The saturation process in \GTO\ is not trivial.
At $T=0.5$~K, instead of a gradual decrease in the gradient of the $M_\parallel(H)$ curve on approaching the saturation field $H_{sat}$, which would be typical for an ordinary antiferromagnet, the gradient increases from a lower field value of $\approx 0.10$ to $\approx 0.13 \mu_B/$kOe per Gd ion at $H_{sat}$.
Although an additional transition at 10~kOe can be seen as a $\sim 10$\% decrease in the slope of the \MlongH\ curve measured at 0.5~K, it becomes more obvious after differentiation.
Fig.~\ref{Fig2_M} suggests that (a) this transition is not particularly temperature dependent up to 0.8~K, and that (b) its influence on the $dM_\parallel/dH(H)$ curves is much more pronounced than the transition at a half of the saturation field.

\begin{figure}[tb]
    \includegraphics[width=0.95\columnwidth]{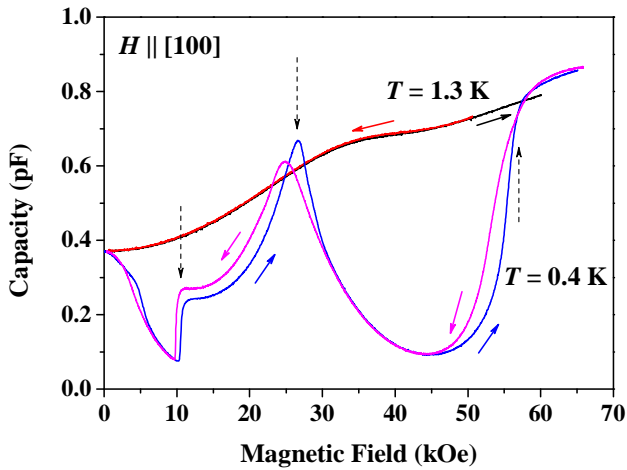}
    \caption{\label{C(H)} (Colour online) Field dependence of the torquemeter capacitance measured at 0.4 and 1.3~K.
    Dashed vertical arrows mark the transition fields. Solid arrows indicate increasing and decreasing field sweeps.}
\end{figure}

The experimentally observed field dependence of the torquemeter capacitances is shown in Fig.~\ref{C(H)}.
The smooth continuous variation of the capacitance with an applied field, as measured at temperatures above the magnetic ordering, is due to a field gradient at the sample position, inhomogeneity in the sample \mH, or a slight sample misalignment causing the de\mH\ field to deviate from the direction of the external field.
All of these effects are proportional to the longitudinal \mH.
Since the variation of \Mlong\ with temperature amounts to only a few percent over the entire range of temperatures explored, one can use the high-temperature ($T>T_{N1}$) response curves as a background for the low-temperature measurements.

On cooling below \Tno\ the torquemeter response curves change drastically.
At the base temperature of the $^3$He cryostat of 0.4~K, the curve has a weak kink around 5~kOe and well defined, abrupt changes around 10, 25, and 55~kOe, presumably corresponding to magnetic field-induced phase transitions.
The hysteresis of the torquemeter response curves at the high-field transition is related to a strong magnetocaloric effect.\cite{Sosin_PRB_2005}
Overall the thermodynamics of GTO is quite complicated near the saturation field, but for temperatures below the magnetic ordering the magnetocaloric effect heats the sample if the field is decreasing and cools the sample if the field is increasing for $H<H_{sat}$.
Since the sample is thermalised only through the 100~$\mu$m-thick bronze cantilever, this may lead to a deviation of the sample temperature from the sensor temperature.
The temperature difference estimated from the spread of the transition points is 25~mK at 0.4~K and up to 100~mK at around 0.7~K.
At low fields (15~kOe and below) any hysteresis in the torquemeter response is not sensitive to the field sweep rate, suggesting that in this field regime the hysteresis is not a thermalization issue.

The transverse \mH\ can be recovered from the experimental capacitance curves as follows.
The magnetic torque is compensated by an elastic force, which is proportional to the change of the capacitor spacing $\Delta d$.
Magnetic torque includes  a transverse \mH\ effect $a M_\perp H$ and a background response $b M_{||}(T>T_{N1}) H$.
For a flat capacitor $d\propto \frac{1}{C}$, hence:
\begin{eqnarray}
\Delta\frac{1}{C} & \propto &(a M_\perp +{b} M_{||}(T>T_{N1}))H \nonumber \\
M_\perp(H,T) & \propto &
\frac{1}{H}\left(\frac{1}{C(H,T)}-\frac{1}{C(H,T>T_{N1})}\right) \nonumber
\end{eqnarray}

\begin{figure}[tb]
    \includegraphics[width=0.95\columnwidth]{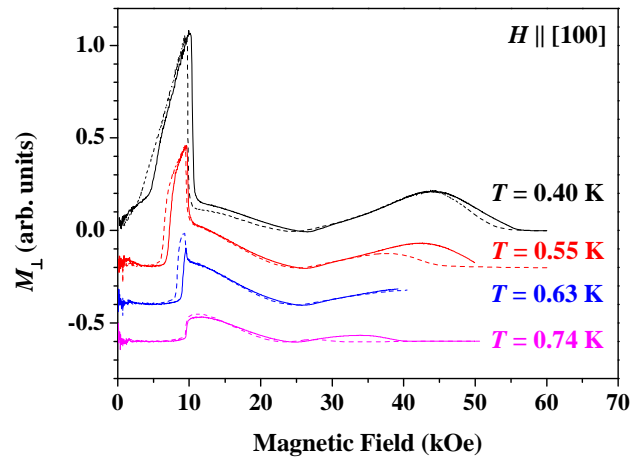}
    \caption{\label{Mtrans(H)} (Colour online) Field dependence of the transverse \mH\ at different temperatures for increasing (solid lines) and decreasing (dashed lines) magnetic fields.
    For clarity, the curves are consecutively offset by -0.2~units of $M_\perp$.}
\end{figure}

The field dependence of the extracted \Mtran\ is shown in Fig.~\ref{Mtrans(H)}.
These curves demonstrate well defined features at the phase transitions.
Firstly, as expected in the paramagnetic (saturated) phase, \Mtran\ is absent above the saturation field of $55\pm1.5$~kOe (here we follow the 0.4~K curve).
\Mtran\ appears below this field and on lowering the field further it vanishes again in a field of $25\pm2$~kOe.
This transition field is close to the $\approx 30$~kOe transition field observed for other studies in which the magnetic field was applied along [111], [110], or [112].\cite{Petrenko_PRB_2004,Glazkov_JPCM}
The transverse moment reappears in lower fields and smoothly increases as the field decreases to $H^{\downarrow}=9.9\pm 0.2$~kOe.
At this field, corresponding to a small change in the slope of the \MlongH\ curve (see Fig.~\ref{Fig2_M}), a sharp step-like increase of the \mH\ component normal to the cantilever plane is observed.
On further decrease of the field the transverse \mH\ decreases almost linearly with the field showing a weak kink below 3~kOe.

On sweeping the field back up at the same temperature of 0.4~K, this kink is transformed into a more extended transition region which terminates with a sharp feature at about 6~kOe.
Presumably this region, marked by a considerable hysteresis, is associated with the reduction of the system into a single magnetic domain state possessing the largest \mH\ component perpendicular to the sample plane.
As the magnetic field is increased further this component decreases sharply by about a factor of 10 at a field $H^{\uparrow}=10.4\pm0.1$~kOe.
The hysteresis effects at this critical field remained largely unchanged when the field sweep rate was reduced by a factor of 3, pointing to the first-order nature of the transition.
This transition is almost temperature independent, while the kink at 5~kOe evolves rapidly with increasing temperature.
At a temperature of 0.55~K, \Mtran\ is zero in low fields, then it appears in a step-like fashion, increases linearly with field up to 10~kOe and then decreases.
At temperatures above 0.6~K, \Mtran\ remains zero in the entire field range below 10~kOe.

\begin{figure}[tb]
    \includegraphics[width=0.95\columnwidth]{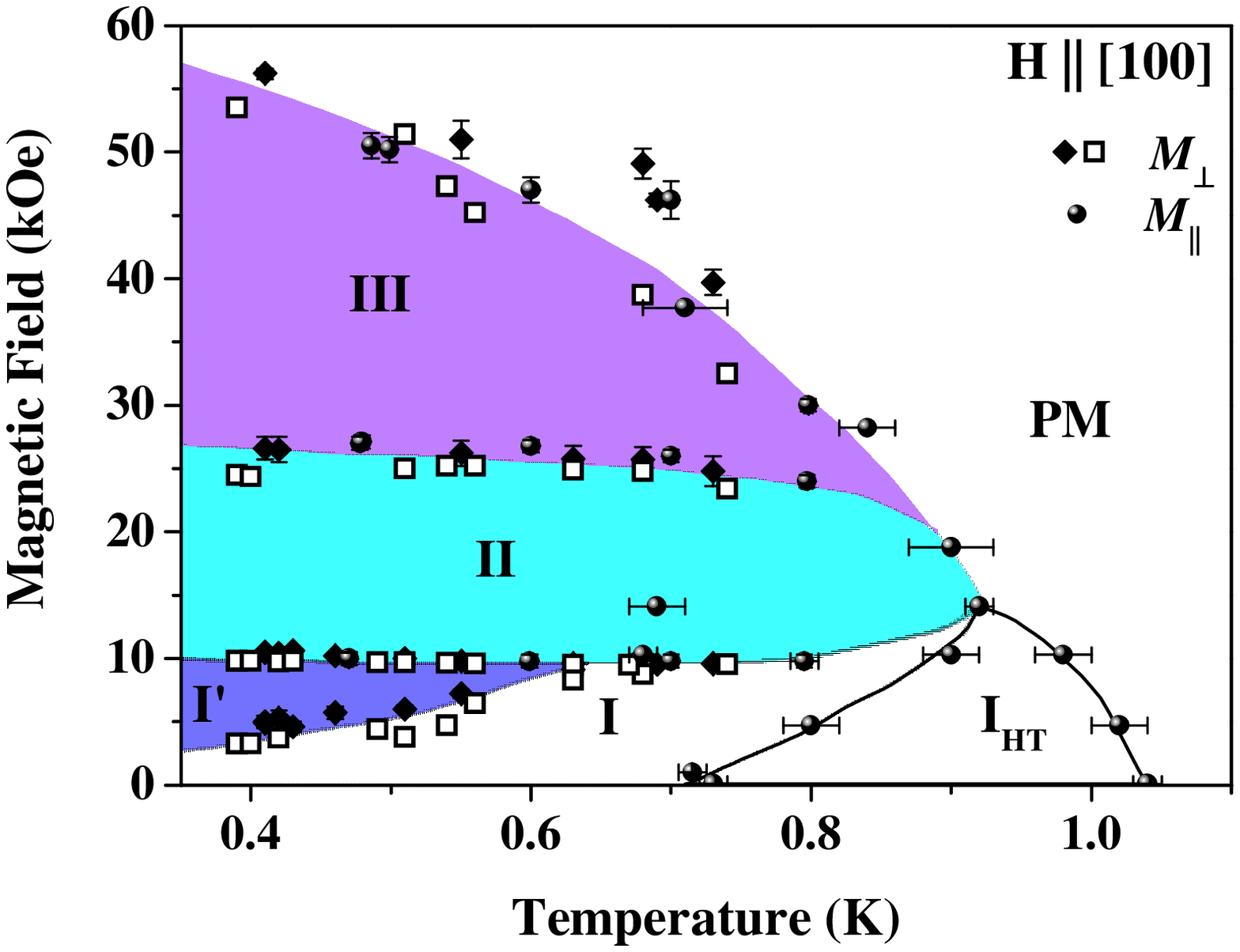}
    \caption{\label{phasediag} (Colour online) Magnetic phase diagram of \GTO\ for a field applied along the [100] axis.
    The phase transition points detected by measuring the field or temperature dependence of the longitudinal \mH\ are marked with circles.
    The open and solid squares were obtained from the transverse \mH\ measurements for decreasing and increasing magnetic fields respectively.
    The lines are guides  to the eyes, colour/white areas on the diagram correspond to the phases with/without transverse \mH.}
\end{figure}

The data presented above, which were collected for a magnetic field applied along the [100] direction, allow one to construct an $H$-$T$ phase diagram which looks distinctly different from those observed for other field orientations (see Fig.~\ref{phasediag}).
We associate phase I in this diagram with the zero-field structure identified in neutron diffraction experiments\cite{Stewart_JPCM_2004} as a partially disordered multi-$k$ structure.
At low-temperature, e.g. $T=0.4$~K, phase I is transformed by weak magnetic fields into phase I$^{\prime}$ in the way that resembles the reduction of a magnetic structure into a single domain state. This process is accompanied by a rapid step-like increase in the transverse component of the \mH\ forbidden by the cubic symmetry of the magnetic structure in phase I.
Observed for \Hloo\ only, this lower symmetry state in which the transverse \mH\ grows linearly in field on the phase diagram is rather unstable.
Heating from within phase  I$^{\prime}$ to above 0.5-0.6~K restores the state I in which the transverse magnetic moment remains {\it exactly} zero for fields below 10~kOe.
Phases I and I$^{\prime}$ are separated from phase II by a first-order transition which was not observed in this field range for any other field directions.
The phase boundary is practically temperature independent and according to our preliminary electron spin resonance measurements\cite{ESR_Sosin_2012} is accompanied by an abrupt change of the gaps in the spin-wave spectrum.
The transverse moment in phase~II (see Fig.~\ref{phasediag}) is either much smaller, but still nonzero in value, or it is rotated in the plane perpendicular to the applied magnetic field.
This phase extending to $\simeq 25$-27~kOe, is also non-cubic in symmetry, and is likely to be equivalent to the low-field states observed in GTO for other field directions.\cite{Petrenko_PRB_2004,Glazkov_JPCM}
The field of 25-27~kOe is similar in value to the transition fields for other directions, but the situation for \Hloo\ is different in that the transition produces no impact on the excitation spectrum,\cite{ESR_Sosin_2012} and does not result in a significant change in the magnetic susceptibility.
This transition from II to III is therefore second-order, unlike the transitions observed for other directions of applied field e.g.  \Hlll.\cite{Sosin_PRB_2006,Petrenko_JPCM_2011}
The high-field part of the phase diagram (phase~III) as well as the higher-temperature phase~$\rm I_{HT}$ at $T_{N2}<T<T_{N1}$ are common for all field directions.

In conclusion, the magnetic phase diagram of \GTO\ for \Hloo\ is established from longitudinal and transverse \mH\ measurements.
At low temperatures and fields, it contains an additional phase with a lower than cubic symmetry, which separates a zero-field magnetic structure from higher-field-induced states.
At higher temperatures this phase disappears and breaking of the cubic symmetry occurs via a first-order transition at a field of about 10~kOe.
A direct identification of the magnetic phases as well as a clarification of the role of the magnetic field in their formation remains a challenge for neutron diffraction experiments.

The Kapitza Institute group acknowledge support from the Russian Foundation for Basic Research (RFBR Grant No.10-02-01105) and the Russian President Program for the Support of the Leading Scientific Schools (Grant No.4889.2012.2).
The Warwick group acknowledges financial support from the EPSRC, UK.
Some of the equipment used at the University of Warwick was obtained through the Science City Advanced Materials project ``Creating and Characterising Next Generation Advanced Material'' with support from Advantage West Midlands and part funded by the European Regional Development Fund.

\end{document}